\documentclass[aps,prd,twocolumn,showpacs,preprintnumbers,footnote,amssymb]{revtex4}
\usepackage{epsfig}
\usepackage{dcolumn}
\usepackage{bm}
\usepackage{amsmath,amssymb}

\begin{document}
\newcommand{\M}{\mathbb{M}}
\newcommand{\R}{\mathbb{R}}
\newcommand{\HI}{\mathbb{H}}
\newcommand{\C}{\mathbb{C}}
\newcommand{\TI}{\mathbb{T}}
\newcommand{\E}{\mathbb{E}}
\def\etal{{\it et al}.} \def\e{{\rm e}} \def\de{\delta}
\def\dd{{\rm d}} \def\ds{\dd s} \def\ep{\epsilon} \def\de{\delta}
\def\goesas{\mathop{\sim}\limits} \def\al{\alpha} \def\vph{\varphi}
\def\Z#1{_{\lower2pt\hbox{$\scriptstyle#1$}}}

\newcommand{\be}{\begin{equation}}
\newcommand{\ee}{\end{equation}}
\newcommand{\bea}{\begin{eqnarray}}
\newcommand{\eea}{\end{eqnarray}}
\newcommand{\nn}{\nonumber}

\def\IR{{\hbox{{\rm I}\kern-.2em\hbox{\rm R}}}}


\renewcommand{\thefootnote}{\fnsymbol{footnote}}
\long\def\@makefntext#1{\parindent 0cm\noindent \hbox to
1em{\hss$^{\@thefnmark}$}#1}
\def\Z#1{_{\lower2pt\hbox{$\scriptstyle#1$}}}


\title{\bf Extra dimensions, warped compactifications and cosmic acceleration}

\preprint{UOC-TP 013/09, CAS-KITPC/ITP-091}

\author{Ishwaree P. Neupane}
\affiliation{Department of Physics and Astronomy, University of
Canterbury, Private Bag 4800, Christchurch 8041, New Zealand, {{\sl
E-mail: ishwaree.neupane@canterbury.ac.nz}}, and \\
Kavli Institute for Theoretical Physics China, CAS, Beijing 100190,
China}

\begin{abstract}

We report on explicit cosmological solutions within the framework
of an inflating de Sitter brane embedded in five- and
ten-dimensional bulk spacetimes. In the specific example we study
the brane tension is induced by the curvature related to the
expansion of a physical $3+1$ spacetime rather than by a bulk
cosmological term. In a generic situation with nonzero brane
tension, the expansion of the universe accelerates eventually. We
also show that inflationary cosmology is possible for a wide class
of metrics without violating four- and higher-dimensional null
energy condition.

\end{abstract}


\maketitle


\section{Introduction}

The one major development that was not anticipated was the
discovery that the expansion of the today's universe is
accelerating~\cite{supernovae}, rather than slowing down. Since an
epoch of cosmic acceleration plays an important role in modern
cosmological models, it would be very interesting to know whether
or not this effect can be understood or explained within the
framework of fundamental theories, including superstring and
supergravity models.

In recent years, several attempts have been made to find explicit
cosmological solutions of ten- and eleven-dimensional supergravity
models that allow accelerating universes using time-dependent
scalar fields or metric
moduli~\cite{TW03,Ohta:2003pu,Maeda:2004hu}. Time-dependent
solutions in pure supergravity generally require some of the extra
spaces to be negatively curved, if they are to allow a cosmic
acceleration of the usual $3+1$ spacetime.

There are a couple of disadvantages of using explicit
time-dependent scalar fields. First, in many examples studied in
the literature, with maximally symmetric extra dimensions, we
usually obtain only a transiently accelerating universe with
time-dependent volume moduli, see
e.g.~\cite{Ohta:2003pu,Ish03a,Ohta:04b}. Second, cosmological
solutions with time-dependent scalar fields usually contain
time-like singularities. This last feature (of a cosmological
solution) is generally unacceptable because generic singularity of
a time-dependent solution in pure supergravity may not have any
quantum interpretation.

In~\cite{Ish05a} it was first realized that cosmological solutions
without any time-like singularities can be obtained by introducing
one or more geometric twists in the extra dimensions which
generate in lower dimensions some nontrivial metric flux. Yet the
corresponding solutions do not lead to a four-dimensional de
Sitter (or quasi de Sitter) spacetime as is required to describe
the inflationary epoch of the universe at its early stages and/or
the present universe with a period of accelerating expansion. We
therefore seek to an alternative scenario with warped extra
dimensions.

In 1999, Randall and Sundrum in a theory referred to as
RS1~\cite{RS1} realized that a five-dimensional braneworld model
with a brane can address the mass hierarchy in particle physics if
there is a second brane some distance away from the first, which
perhaps mimics the observed $3+1$ spacetime. An even more
revolutionary idea was that gravity can be `trapped' (on a brane)
and extra dimensions may have infinite spatial extent~\cite{RS2}.
For this simple and elegant proposal to work one needs a
five-dimensional anti de Sitter space, i.e. a background geometry
which is negatively curved, which suppresses the effect of warping
at the brane's position or the 4D hypersurface, leading to a zero
cosmological constant. Once the bulk cosmological term is assumed
to be zero then the RS solution would be lost. It is therefore of
natural importance (and our interest) to find nontrivial
(cosmological) solutions that exist in flat spacetimes as well.

One simple thing that can happen when we view our observed
universe as a cosmological brane embedded in a higher-dimensional
spacetime is that the universe can accelerate because of an
effective four-dimensional cosmological constant induced on the
brane or due to the warping of additional spatial dimensions. In
general, this phenomenon can (and should perhaps) occur when the
tension on the brane(s) is positive.

Following~\cite{RS1,RS2}, we find interest in warped metrics that
maintain the usual four-dimensional Poincar\'e symmetry, with
general metric parametrization:
\begin{equation}\label{10d-metric-gen}
ds_{D}^2 = W(y)^2\, \hat{g}_{\mu\nu} dX^\mu dX^\nu + W(y)^{\gamma}
{g}_{mn}(y)\, dy^m dy^n,
\end{equation}
where $X^\mu$ are the usual spacetime coordinates ($\mu, \nu =0,
1, 2, 3$), $W(y)$ is the warp factor as a function of one of the
internal coordinates and $\gamma$ is a constant. Non-factorizable
metrics as above can be phenomenologically motivated as in
five-dimensional braneworld models as well as in ten- and
eleven-dimensional supergravity models with reduced
(super)symmetries~\cite{Gibbons-84,deWit86a,Malda-Nunez}. They can
also arise naturally in string theory compactification with
flux~\cite{Polchinski:1995,KMBecker,GKP,KKLT}.

In this Letter by considering the metric~(\ref{10d-metric-gen}),
we present explicit cosmological solutions for which not only the
warp factor is nontrivial but also the physical $3+1$ spacetime
undergoes an inflationary de Sitter expansion, especially, when
the brane tension is nonzero. An intriguing feature of such new
solutions is that the scale factor of the universe becomes a
constant only in the limit where the warp factor $W(y)$ also
becomes a constant. In a sense, the warp factor cannot be a
constant except in the region where the scale factor of the
universe is also constant, leading to a Minkowski spacetime.

For generality, we take the full spacetime dimensions to be $D$,
which we split as $D\equiv 4+m \equiv 4+1+q$. The internal
$m$-dimensional manifold is assumed to be an Einstein space
$$
ds\Z{D-4}^2= {g}_{mn}(y)\, dy^m dy^n
$$
having positive, negative or zero Ricci scalar curvature
($R^{(m)}>0$, $R^{(m)}<0$ or $R^{(m)}=0$). We should note that the
choices made by Gibbons~\cite{Gibbons-84}, Maldacena and
Nunez~\cite{Malda-Nunez} and Giddings et al.~\cite{GKP}, with
respect to warped compactifications, are all different. These are,
respectively, $\gamma=0$, $\gamma=2$ and $\gamma=-2$. This
difference may not be much relevant in $D=5$ dimensions: the
reason being that an arbitrary metric $g\Z{55}(y)$ times an
arbitrary power of the warp factor is still an arbitrary metric.
However, in dimensions $D\ge 6$, the choice of $\gamma$ would be
relevant since it ought to be related to the Ricci curvature of
the internal manifold as explicitly shown in~\cite{Ish:09a}; we
just need to relate the coefficient $\gamma$ to $R^{(m)}$.
Especially, for the discussion of no-go theorems
in~\cite{Gibbons-84,Malda-Nunez,GKP}, the choice $\gamma$ is not
very important, for the theorems of these papers ruled out the
existence of de Sitter solutions in pure supergravity just because
of an extra condition on the warp factor, so-called the
boundedness condition $\int \nabla^2 W^4 = \int
\left(W^4\right)^{\prime\prime}-2\gamma \int W^2 {W^\prime}^2=0$,
which is, however, not satisfied by cosmological solutions,
especially, when the extra dimensions are only geometrically
compact and/or when there are localized sources like branes and
orientifold planes. In our analysis below we shall relax the
condition like $\int \nabla^2 W^n=0$ (where $n$ is some constant)
until we are ready to comment on this part of the problem.

One could naively think that the coefficient $\gamma$ plays no
role in the discussion of warped compactifications. The reason is
that, since the metric ${g}_{mn}(y)$ is arbitrary, an arbitrary
metric times an arbitrary power of the warp factor is still an
arbitrary metric. Here one should also note that the metric
$g_{mn}(y)$ is {\it not} just a single canonical function of $y$
but has more than one components, $(m,n)=1, 2, \cdots (D-4)$. In
dimensions $D\ge 6$, one cannot absorb $W^{\gamma}$ into
$g_{mn}(y)$ just by using some coordinate transformations unless
that each and every components of $g_{mn}(y)$ are equal or
proportional to the same function, say $f(y)$. For clarity, let us
take $D=10$ and write the 6d metric as
\begin{equation}
ds\Z{6}^2 = h(y)\,dy^2 + f(y)\, \tilde{g}_{mn} d\Theta^m
d\Theta^n,
\end{equation}
where $\tilde{g}_{mn}$ denote the metric components of the
five-dimensional base space $X_5$, which are independent of the
$y$ coordinate. The volume factor $W^\gamma$ in
Eq.~(\ref{10d-metric-gen}) may be absorbed inside $dy^2$ by using
the transformation $W^\gamma h(y) dy^2 \equiv d\tilde{y}^2$ and
also defining a new function $X(\tilde{y})$ such that $W^\gamma
f(y) \equiv X(\tilde{y})$. With these substitutions, the warp
factor $W^2$ multiplying the 4d part of the metric is
$[X(\tilde{y})/f(\tilde{y})]^{2/\gamma}$. The 10D metric still
involves two unknown functions and the free parameter $\gamma$.
That is to say, if we want to write a general metric ansatz (for
the purpose of solving Einstein's equations), then we have to
allow one more free parameter in the metric than that were
considered in~\cite{Gibbons-84,Malda-Nunez,GKP}.

It is not difficult to check that only a specific value of
$\gamma$ would give a nontrivial cosmological solution, once we
specify the 6d metric or fix the spatial curvature of the internal
space. Suppose we chose $\gamma=0$ and then simultaneously assumed
that the internal space is Ricci flat, then we would not find a de
Sitter solution at least in pure supergravity. The story would be
similar for some other specific choices of $\gamma$ and/or the
internal curvature. For example, if we set $\gamma=-2$ in
(\ref{10d-metric-gen}), then we would find a de Sitter solution
only by allowing $Y_6$ to have negative curvature. In view of this
discussion, at this stage we shall keep both the coefficient
$\gamma$ and the curvature of the internal space arbitrary.

\section{An explicit model in $D=5$ dimensions}

Let us first consider a specific example where the real world
looks like a five-dimensional universe described by the metric
\begin{equation}\label{5dmetric-gen}
ds\Z{5}^2 = W(y)^2\, \hat{g}_{\mu\nu} dX^\mu dX^\nu +
\rho^2\,W(y)^{\gamma}\,dy^2,
\end{equation}
where $\rho$ is the radius of compactification, which may be
assumed to be a constant in the simplest scenario under
consideration. The classical action describing this warped
geometry is given by
\begin{equation}\label{main-action1}
S= \frac{M\Z{5}^3}{2}  \int d^5 {x}  \sqrt{-g\Z{5}}  \,R_{(5)},
\end{equation}
where $M_5$ is the fundamental 5D Planck scale. Our starting point
is different from that in the RS braneworld models only in that we
take the metric of the usual four-dimensional spacetime in a
general form
\begin{eqnarray}\label{FRW}
ds_4^2 &=&   -dt^2+  a^2(t)\left[\frac{ dr^2}{1-k r^2}+ r^2
(d\theta^2+\sin^2\theta d\phi^2)\right]\nonumber \\
&\equiv &  \hat{g}_{\mu\nu} dX^\mu dX^\nu,
\end{eqnarray}
where $a(t)$ is the scale factor of the universe. We have allowed
all three possibilities for the physical 3D spatial curvature:
flat ($k=0$), open ($k<0$) and closed ($k>0$). Models similar to
the one here were studied before, see for
example~\cite{Kaloper:99a,Himemoto00b}, but an interesting (and
perhaps new) observation is that for the existence of inflationary
de Sitter solutions we do not necessarily require a 5D bulk
cosmological constant term.

The three independent Einstein's equations following from the
metrics (\ref{5dmetric-gen}) and (\ref{FRW}) are given by
\begin{subequations}
\begin{align}
{W^\prime}^2- \rho^2
\left(\frac{\dot{a}^2}{a^2}+\frac{k}{a^2}\right)W^\gamma
 =0,\label{sf-part1} \\
2 W W^{\prime\prime} - \gamma {W^\prime}^2=0,\label{sf-part2}\\
\frac{\ddot{a}}{a} -\frac{\dot{a}^2}{a^2}-\frac{k}{a^2}
=0,\label{sf-part3}
\end{align}
\end{subequations}
where ${}^\cdot$ and ${}^\prime$ denote respectively
$\partial/\partial{t}$ and $\partial/\partial{y}$ (or
$\partial/\partial z$ when $\gamma=2$). From Eq.~(\ref{sf-part3})
we immediately obtain
\begin{equation}\label{main-sol-scale1}
a(t)= \frac{1}{2} \exp\left(\frac{\mu (t-t\Z{0})}{\rho} \right)
+\frac{k\,\rho^2}{2\mu^2}\, \exp\left(\frac{\mu
(t\Z{0}-t)}{\rho}\right),
\end{equation}
where $\mu$ and $t\Z{0}$ are integration constants. In the
$\gamma\ne 2$ case, from Eqs.~(\ref{sf-part1}) and
(\ref{sf-part2}), we obtain
\begin{equation}\label{sol-5Db}
\left[W(y)\right]^{2-\gamma}  = \frac{1}{4}
\left(2-\gamma\right)^2 \mu^2 (y+c)^2.
\end{equation}
The bulk singularity at $y=-c$ is just a coordinate artifact,
which could simply be absent in some other coordinate systems. To
quantify this we can either introduce a new coordinate $z$
satisfying $W(y)^{\gamma/2} dy \equiv W(z)\, dz$ or solve the 5D
Einstein equations by setting $\gamma=2$ in (\ref{5dmetric-gen})
and replacing $y$ there by $z$. We then get
\begin{equation}\label{sol-5Da}
ds_5^2 = e^{- 2 \mu z} \left(\hat{g}_{\mu\nu} dX^\mu dX^\nu +
\rho^2\, {dz}^2\right).
\end{equation}
One could in principle set $\rho=1$ in Eq.~(\ref{sol-5Da}) or in
Eq.~(\ref{main-sol-scale1}), but an essential point here is that
the scale factor and warp factor can have quite different slopes.
In natural Planck's unit one may require $\rho\ll 1$ (see below).
Note that, with $\mu>0$, the universe must accelerate eventually.
In a sense the universe accelerates due to a kind of back reaction
of the 5D warped geometry on the usual four-dimensional spacetime.


For the above solution the 4D effective Newton's constant is not
finite. The reason being that in (\ref{sol-5Da}) $z$ ranges from
$-\infty$ to $+\infty$, and hence the extra dimension has infinite
warped volume. In order to get physical results, including a
finite 4D Newton's constant, we shall introduce some elements of
RS type braneworld models.

\subsection{A geometrically compact extra dimension}

To this end, we specify a boundary condition such that the warp
factor is regular at $z=0$ where we place a 3-brane with brane
tension $T_3$. We also introduce a bulk cosmological term
$\Lambda$. The classical action describing this set up is
\begin{eqnarray}
S= \frac{M\Z{5}^3}{2} \int d^5{x} \sqrt{-g\Z{5}} \left(R
-2\Lambda\right)+ \frac{M\Z{5}^3}{2} \int d^4 x\sqrt{-g_{\rm
b}}(- T_3), \label{main-action2a}\nonumber \\
\end{eqnarray}
where $g_{b}$ is the determinant of the metric $g_{ab}$ evaluated
at $z=0$. Einstein's equations are given by
\begin{eqnarray}
G_{AB} =- \frac{T_3}{2}\, \frac{\sqrt{-g\Z{b}}}{\sqrt{-g}}
g_{\mu\nu}^{b} \delta_A^\mu \delta_B^\nu \delta(z) -\Lambda
g_{AB}.
\end{eqnarray}
Eqs.~(\ref{sf-part1}) and (\ref{sf-part2}) get modified as
\begin{subequations}
\begin{align}
{W^\prime}^2- \rho^2\left(
\frac{\dot{a}^2}{a^2}+\frac{k}{a^2}\right) W^\gamma
=- \frac{\hat{\Lambda}}{6} W^{2+\gamma}, \label{main-first-eq}\\
2 W W^{\prime\prime} - \gamma {W^\prime}^2=-
\frac{\hat{\Lambda}}{3}W^{2+\gamma} - \frac{\tau\Z{3}\,
\delta(z)}{3}W^{2-\gamma/2},\label{main-second-eq}
\end{align}
\end{subequations}
where $\hat{\Lambda}\equiv \Lambda \,\rho^2$ and $\tau\Z{3}\equiv
T_3 \,\rho^2$, while Eq.~(\ref{sf-part3}) is the same, which is
unaffected by a bulk cosmological term. One replaces $\delta(z)$
by $\delta(y-y\Z{0})$ in the $\gamma\ne 2$ case. With the widely
used choice that $\gamma=0$, we get~\cite{Garriga:1999}
\begin{equation}
W(y)= \frac{\sqrt{6}}{\mu\sqrt{-\hat{\Lambda}}}\sinh
\left[\frac{\sqrt{-\hat{\Lambda}}\, (y+c)}{\sqrt{6}}
\right].\label{sol-gamma0}
\end{equation}
By defining $W(y)^{\gamma/2} dy \equiv W(z)\, dz$, we obtain
\begin{equation}\label{main-sol-5D}
W(z)=\frac{24 \mu^2}{24 \mu^2 e^{\mu |z|}+ \Lambda \rho^2\,
e^{-\mu |z|}}\,,
\end{equation}
which has a smooth $\Lambda\to 0$ limit. This result is nothing
but an exact solution of 5D Einstein equations with $\gamma=2$ in
Eq.~(\ref{5dmetric-gen}). In the above we demanded a $Z_2$
symmetry about the brane's position at $z=0$. If we relax this
symmetry, then the warp factor becomes singular at
$z=-\frac{1}{2\mu} \ln
\left(-\frac{24\mu^2}{\Lambda\rho^2}\right)$, especially, with
$\Lambda<0$. We shall therefore consistently demand a $Z_2$
symmetry about the brane's position at $z=0$, irrespective of the
choice $\Lambda=0$ or $\Lambda<0$. It is not difficult to check
that Einstein's equations are satisfied at $z=0$ when
\begin{equation}
T_3=\frac{24\mu^2-\Lambda \rho^2}{2\mu\rho^2}.
\end{equation}
The brane tension is positive when $\mu^2 > - \Lambda \rho^2/24$.
As in RS models~\cite{RS1,RS2}, the choice $\Lambda<0$ could be
more physical.

The solution (\ref{main-sol-5D}) is defined up to a rescaling of
$z$ coordinate, implying that
\begin{equation}
W(z)=\frac{24 \mu^2}{24 \mu^2\, e^{\mu (|z|+z_0)} + \Lambda
\rho^2\, e^{- \mu (|z|+z_0)}}.
\end{equation}
In the $\Lambda=0$ case, we take $z_0=0$ so that $W(z)=1$ at
$z=0$. In the $\Lambda< 0$ case, we take
\begin{equation}
e^{\mu z_0}=\frac{1}{2} + \frac{1}{2}
\sqrt{1-\frac{\Lambda\rho^2}{6\mu^2}}.
\end{equation}
In the limit $\mu\to 0$, we get $W(z)\to 1$ and $a(t)\to {\rm
const}$ (especially when $k=0$), giving rise to a 5D Minkowski or
AdS$_5$ spacetime depending on the choice that $\Lambda=0$ or
$\Lambda<0$.

The four-dimensional effective theory follows by substituting
Eq.~(\ref{5dmetric-gen}) into the classical
action~(\ref{main-action2a}). Here we focus on the 5D curvature
term from which we can derive the scale of gravitational
interactions:
\begin{eqnarray}
S_{\rm eff} & \supset & \frac{M_5^3\,\rho}{2} \int d^4 x
\sqrt{-\hat{g}\Z{4}} \int  dz\, W^{2+\gamma/2}\nonumber\\
&{}& \qquad \times \left(\hat{R}_4 -{\cal L}\Z{0}-2\Lambda W^2
\right),
\end{eqnarray}
where ${\cal L}\Z{0} \equiv \rho^{-2} \,W^{-\gamma} \left(12
{W^\prime}^2
 + 8 W W^{\prime\prime} -4\gamma {W^\prime}^2 \right)$.
 As a simple example, henceforth we take $\gamma=2$. Hence
\begin{eqnarray}
{\cal L}_0 &=& \frac{12\mu^2} {\rho^2} \left(1-\frac{160\Lambda
\mu^2 \rho^2}{(24\mu^2\,e^{\mu |z|}+\Lambda \rho^2\,e^{-\mu
|z|})^2}\right) \nonumber \\
&{}& \quad - \frac{16\mu}{\rho^2} \left(\frac{24\mu^2 e^{\mu
|z|}-\Lambda \rho^2 e^{-\mu |z|}}{24\mu^2 e^{\mu |z|}+\Lambda
\rho^2 e^{-\mu |z|}}\right)\delta(z).
\end{eqnarray}
This result shows that a negative bulk cosmological term could
make the value of 4D effective cosmological constant more
positive. In the particular case that $\Lambda=0$, the above
expression takes a much simpler form
\begin{equation}
{\cal L}\Z{0}= \rho^{-2} \left(
 12\mu^2 - 16\mu \delta(z)\right).
\end{equation}
The relation between four- and five-dimensional
 effective Planck masses is then given by
\begin{equation}
 M_{\rm Pl}^2 = {M_5^3\,\rho}\int_{-\infty }^{\infty} d{z}\,e^{-3\mu |z|}=
 \frac{2 M_5^3 \rho}{3\mu}.
\end{equation}
In the limit $\mu\to 0$, the extra dimension $z$ opens up and we
thus obtain a 5D Minkowski space ($M_{\rm Pl}^2\to \infty$ or
$G_4\to 0$). However, in the generic situation with $\mu>0$, the
4D Newton's constant is finite.

Although the details and the motivations are different, the
$\Lambda=0$ solution above bears certain features of a 5D
braneworld model discussed by Dvali et al.~\cite{Dvali-etal} where
a cosmic (self-)acceleration of the universe is supported by the
4D scalar curvature term on the brane. In the present approach,
however, the 5D spacetime is non-factorizable and the universe
accelerates because of a positive curvature ($R_4>0$)
induced by the 5D warped geometry. 

\subsection{A physically compact extra dimension}

The above analysis can easily be extended to a set up with two
3-branes, as in RS1 braneworld model. To this end, one introduces
a 5D bulk cosmological term $\Lambda$ and also specifies boundary
conditions such that the warp factor is regular both at orbifold
fixed points $y=0$ and $y=\pi $ where we place two 3-branes ($b_1$
and $b_2$) with brane tension $T_3^{(1)}$ and $T_3^{(2)}$,
respectively. We start with a canonical metric (choosing
$\gamma=0$ in~(\ref{5dmetric-gen}))
\begin{equation}
ds\Z{5}^2 = W(y)^2\, \hat{g}_{\mu\nu} dX^\mu dX^\nu +
\rho^2\,dy^2,
\end{equation}
where as above $0\le y \le \pi$ is the coordinate for an extra
dimension, which is a finite interval whose size is set by $\rho$.

The classical action describing this set up is
\begin{eqnarray}
S &=& \frac{M\Z{5}^3}{2}\Big( \int d^5{x} \sqrt{-g\Z{5}} \left(R
-2\Lambda\right)+  \int d^4 x\sqrt{-g\Z{\text b1}}(- T_3^{(1)})
\nonumber\\
&{}& \quad + \int d^4 x\sqrt{-g\Z{\text b2}}(- T_3^{(2)})\Big),
\label{main-action2}
\end{eqnarray}
where $g_{b1}$ and $g_{b2}$ are determinants of the metric
$g_{ab}$ evaluated at $y=\pi$ and $y=0$. The 5D Einstein equations
read
\begin{subequations}
\begin{align}
\frac{\ddot{a}}{a} -\frac{\dot{a}^2}{a^2}-\frac{k}{a^2}
=0,\label{first-eqn} \\
\frac{{W^\prime}^2}{W^2}-
\frac{\rho^2}{W^2} \left(
\frac{\dot{a}^2}{a^2}+\frac{k}{a^2}\right)
+ \frac{\Lambda\rho^2}{6}=0, \label{second-eqn}\\
\frac{ W^{\prime\prime}}{W} + \frac{\rho T_3^{(1)}}{6}
\,\delta(y-\pi) + \frac{\rho T_3^{(2)}}{6}\,\delta(y)+
\frac{\Lambda\rho^2}{6}=0,\label{third-eqn}
\end{align}
\end{subequations}
The solution to Eqs. (\ref{first-eqn})-(\ref{second-eqn})
consistent with the orbifold symmetry $y\to -y$ is
\begin{subequations}
\begin{align}
a(t)= \frac{1}{2} \exp\left(\frac{\mu t}{\rho} \right)
+\frac{k\,\rho^2}{2\mu^2}\, \exp\left(-\frac{\mu
t}{\rho}\right),\\
W(y)= \frac{\sqrt{6}}{\mu\,\sqrt{-\Lambda\rho^2}} \sinh
\left[\frac{\sqrt{-\Lambda\rho^2}}{\sqrt{6}}\,(|y|-y_0)
\right].\label{sol-gamma-new}
\end{align}
\end{subequations}
Note that, in computing derivatives of $W$, we are to consider the
metric a periodic function in $y$. Eq. (\ref{sol-gamma-new}),
valid for $-\pi \le y \le \pi$, then implies
\begin{eqnarray}
&& \frac{W^{\prime\prime}}{W}+\frac{\Lambda \rho^2}{6} +
\sqrt{\frac{-\Lambda \rho^2}{6}}\coth \left(\sqrt{\frac{-\Lambda
\rho^2}{6}}(|y|-y_0)\right)\nonumber \\
&&\qquad \times
\left[2\delta(y-\pi)-2\delta(y)\right]=0.\label{2devts-W}
\end{eqnarray}
Note that, unlike in RS1 brane world model, we do not necessarily
require $T_3^{(1)}=-T_3^{(2)}$; the brane tensions could well
depend on their positions. By placing them at $y=\pi$ and $y=0$,
from Eqs. (\ref{third-eqn}) and (\ref{2devts-W}) we find
\begin{subequations}
\begin{align}
T_3^{(1)}= 2\sqrt{-6\Lambda \rho^2}\coth
\left(\sqrt{\frac{-\Lambda \rho^2}{6}}(\pi-y_0)\right), \\
T_3^{(2)}=-\, 2\sqrt{-6\Lambda \rho^2}\coth
\left(\sqrt{\frac{-\Lambda \rho^2}{6}}(-\,y_0)\right).
\end{align}
\end{subequations}
By defining $\Lambda\equiv -6/L^2$, where $L$ is the curvature
length associated with AdS$_5$ space, we get
\begin{equation}
W(y)=\frac{L}{\rho\mu } \sinh\left(\frac{\rho}{L}
(|y|-y_0)\right).
\end{equation}
The bulk singularity at $|y|=y_0$ may be avoided by taking
$y_0<0$, in which case one of the 3-branes would have a negative
tension. The Goldberger and Wise mechanism to stabilize the size
of fifth dimension or radion using a nontrivial bulk scalar
field~\cite{Goldberger99} may be applied to the present model, but
in this Letter we do not study such effect.


\section{Revisiting braneworld no-go theorems}

The no-go theorems of \cite{Gibbons-84,deWit86a,Malda-Nunez} claim
that vacuum solutions of the type presented above should not
exist, while we have explicitly shown the existence of a
four-dimensional de Sitter solution within 5D Einstein gravity.
There arises an important question as: What prevented the previous
authors from inventing (or ruling out) the explicit de Sitter
solutions given above? To answer this question we need to
carefully examine the conditions embedded in the discussion of the
earlier no-go theorems. Below we will focus on the case of a 5D
Minkowski bulk, but its generalization in higher dimensions should
be straightforward.

\subsection{No-go theorem in five-dimensions}

For the metric (\ref{5dmetric-gen}), the basic equations reduce to
\begin{subequations}
\begin{align}
{}^{(5)}R_{\mu\nu} = {}^{(4)}R_{\mu\nu} -\frac{\hat{g}_{\mu\nu}}{4
W^{\gamma}}\left[ \frac{\left(W^4\right)^{\prime\prime}}{W^2}
-2\gamma
{W^\prime}^2 \right], \\
R_{55} = - \frac{4}{W} W^{\prime\prime} + \frac{2\gamma}{W^2}
{W^\prime}^2.
\end{align}
\end{subequations}
Here, for simplicity, we have set $\rho=1$. We may rewrite the
above two equations as follows
\begin{subequations}
\begin{align}
R_g = R_{\hat{g}} \, W^{-2} - 2(6-\gamma) {W^\prime}^2
W^{-2-\gamma}
-4 W^{\prime\prime} W^{-1-\gamma},\label{mu-mu-part} \\
R_5\,^5 = -4 W^{\prime\prime} W^{-1-\gamma} +2\gamma {W^\prime}^2
W^{-2-\gamma},\label{55-part}
\end{align}
\end{subequations}
where $R_g \equiv {}^{(5)}R_\mu\,^\mu$ and $R_{\hat{g}} \equiv
{}^{(4)}R_\mu\,^\mu$ are, respectively, the curvature scalars of the
5- and 4-dimensional spacetimes with the metric tensors $g_{\mu\nu}$
and $\hat{g}_{\mu\nu}$.
A linear combination of $(1-n) W^{n+\gamma}\times $
Eq.~(\ref{mu-mu-part}) and $(n-4) W^{n+\gamma}\times $
Eq.~(\ref{55-part}) gives (where $n$ is an arbitrary constant)
\begin{eqnarray}\label{consistency1}
&& \frac{(W^n)^{\prime\prime}}{n} -\frac{\gamma}{2} {W^\prime}^2
W^{n-2} \nonumber \\
&&  = \left[\frac{1-n}{12} \left(R_g - R_{\hat g}
W^{-2}\right)+\frac{n-4}{12} R_5\,^5\right]W^{n+\gamma}.
\end{eqnarray}
From the 5D Einstein equations
$$R_{A}^B=8\pi G_5
\left(T_A^B-\frac{1}{3} \,\delta_A^B T_C\,^C\right),$$ we obtain
$${}^{(5)}R_\mu^\mu=8\pi G \left(-\frac{1}{3}\, {}^{(5)}T_\mu^\mu -
\frac{4}{3}\, T_5^5\right)$$ and $$R_5^5= 8\pi G
\left(-\frac{1}{3}\, {}^{(5)}T_\mu^\mu + \frac{2}{3}\,
T_5^5\right).$$ From Eq.~(\ref{consistency1}) we then find
\begin{eqnarray}\label{final-GKK}
&& \left(A^\prime e^{nA}\right)^\prime - \frac{\gamma}{2}
{A^\prime}^2 e^{nA} + \frac{1-n}{12}\, R_{\hat{g}}\,
e^{(n+\gamma-2)A} \nonumber\\
&& \quad  = \frac{2\pi G_5}{3} \left(T_g + (2n-4)
T_5\,^5\right)e^{(n+\gamma)A},
\end{eqnarray}
where $e^{A(y)}\equiv W(y)$ and $T_g\equiv {}^{(5)} T_\mu\,^\mu$.
With $\gamma=0$, we recover the braneworld sum rule discussed
in~\cite{GKK}.

We argue that the warp factor constraints such as $\oint \nabla^2
W^4 =0$ and $\oint \nabla (W^{n-1} \nabla W)=0$ discussed
in~\cite{Gibbons-84,Leblond:2001xr} are `strict', which are not
essentially satisfied by cosmological solutions, especially, when
the extra dimensions are only geometrically compact. For clarity,
take $\gamma=2$ and thus $W(z)=e^{A(z)}= e^{-\mu |z|}$. We then
find
\begin{eqnarray}
\oint \nabla^2 W^4 &\equiv& \oint
\left(W^4\right)^{\prime\prime}-2\gamma \oint {W^\prime}^2
W^2\nonumber \\
&=& 4 \oint W^3 W^{\prime\prime} + 8 \oint W^2 {W^\prime}^2
\nonumber \\
&=&  \oint e^{-4\mu |z|}\left(12\mu^2 - 8\mu \,\delta(z)\right)\ne
0.
\end{eqnarray}
There can be an additional condition on the warp factor, i.e. the
finiteness of 5D warped volume $ 1/G_N\sim \int W^{2+\gamma/2}
=$const. This holds in the above example because of a $Z_2$
symmetry under $z\to -z$.

Coming back to Eq.~(\ref{final-GKK}), and following~\cite{GKK},
let us assume that there exists a class of solutions for which
$\oint \left(A^\prime e^{nA}\right)^\prime=0$, which is plausible
if the extra dimension is like a closed cycle or compact. We then
find
\begin{eqnarray}
 && \oint \left(T_g + (2n-4) T_5\,^5\right)
  e^{(n+\gamma)A}\nonumber\\
  &&\quad  = \frac{1-n}{8\pi G_5} R_{\hat{g}}\oint e^{(n+\gamma-2)A}
-\frac{3\gamma}{4\pi G_5} \oint {A^\prime}^2 \e^{nA}.\nonumber
\end{eqnarray}
We can get $R_{\hat{g}}>0$ by appropriately choosing $n$ or
$\gamma$, even if the term on the left-hand side vanishes. This
result is consistent with some explicit de Sitter solutions of 5D
Einstein equations presented above (cf.
Eqs.~(\ref{main-sol-scale1})-(\ref{sol-5Da})).

In the presence of a bulk cosmological constant $\Lambda$, the 5D
energy momentum tensor is given by
\begin{equation}
T_{AB}=-\frac{1}{8\pi G_5} \left(\Lambda g_{AB} + \frac{T_3}{2}\,
\delta(z) P(g_{AB})\right),
\end{equation}
where $P(g_{AB})\equiv g_{\mu\nu} \delta_A^\mu
\delta_B^\nu/\sqrt{g_{zz}}$. From this we derive
\begin{equation}
 T_g + (2n-4) T_5\,^5= \frac{- 2n
\Lambda -2 T_3 \delta(z)}{8\pi G_5}.
\end{equation}
By demanding that $\Lambda<0$, and with a suitable choice of $n$
or $\gamma$, we can obtain a de Sitter solution, i.e.
$R_{\hat{g}}>0$ even if the brane tension is positive. In the case
$T_3<0$, the cosmic acceleration of a four-dimensional universe
seems more plausible due to an explicit violation of 4D strong
energy condition, but the choice $T_3<0$ is not well motivated (at
least in a single brane set up).

In summary, our results above show that if we do not enforce the
warp factor constraint such as $\int \nabla^2 W^{4}=0$, which does
not hold in several examples considered in this Letter, then it is
possible to realize a cosmological de Sitter solution even within
some simplest or canonical warped braneworld and supergravity
models.

\subsection{No-go theorem in ten-dimensions}

In spacetime dimensions $D\ge 6$ (or $m\ge 2$), with a judicious
choice of $\gamma$, we can find de Sitter solutions with all three
different choices of the internal curvature, i.e. $R^{(m)}=0$,
$R^{(m)}>0$ and $R^{(m)}<0$. This could again be seen in contrast
to the no-go theorems discussed in~\cite{Gibbons-84,Malda-Nunez}.
We should therefore have a closer look on the earlier no-go
arguments. Assuming that ten-dimensional supergravity is the
relevant framework, we may write the 10D metric as
\begin{equation}
ds\Z{10}^2= e^{2A(y)} ds\Z{4}^2 + \rho^2 \,e^{\gamma A(y)}
ds\Z{6}^2\label{10D-nogo}
\end{equation}
with
\begin{subequations}
\begin{align}
ds\Z{6}^2= dy^2+ dy_1^2+ \cdots + dy_5^2,\quad
(\epsilon=0)\label{6d-flat}\\
ds\Z{6}^2= dy^2+ \sin^2{y}
\,d\Omega_{5}^2,\quad (\epsilon=+1) \label{6d-positive}\\
ds\Z{6}^2 =dy^2+ \sinh^2{y}\, d\Omega_{5}^2,\quad (\epsilon=-1),
\label{6d-negative}
\end{align}
\end{subequations}
where $d\Omega_{5}^2$ represents the metric of a usual 5-sphere.
In the above example, the internal 6d manifold is maximally
symmetric, $\tilde{R}_{mn}=\epsilon (m-1) \tilde{g}_{mn}$. A
straightforward calculation gives
\begin{equation}
{}^{(10)}R_{\mu\nu} =
{}^{(4)}\hat{R}_{\mu\nu}-\frac{\hat{g}_{\mu\nu}}{\rho^2} \left(
\nabla_y^2 A + 2 (2+\gamma) {A^\prime}^2\right)
e^{(2-\gamma)A},\label{10Dand4Dcurv}
\end{equation}
\begin{eqnarray}
{}^{(10)}R_{mn}(x,y)&=& {}^{(6)}\tilde{R}_{mn} -2(2+\gamma)
{A^\prime}^2 \tilde{g}_{mn}^{(6)} \nonumber \\
&{}& -\frac{\gamma}{2}\tilde{g}_{mn}^{(6)} \nabla_y^2 A -
2(3+\gamma)\nabla_m \partial_n A\nonumber \\
&{}&  -\left(8-(2+\gamma)^2\right)
\partial_m A \partial_n A,
\end{eqnarray}
where
\begin{equation}
\nabla_y^2 A = \left\{\begin{array}{l} A^{\prime\prime}, \\
A^{\prime\prime}+ 10 A^\prime \cot{y}, \\
A^{\prime\prime}+ 10 A^\prime \coth{y},
\end{array} \right.
\end{equation}
respectively, for the metrics (\ref{6d-flat}), (\ref{6d-positive})
and (\ref{6d-negative}). In the particular case that $\gamma=2$,
we get~\cite{Malda-Nunez}
\begin{eqnarray}\label{10Dcurv}
{}^{(10)}R_{\mu\nu} &=& {}^{(4)}\hat{R}_{\mu\nu} -
\frac{\hat{g}_{\mu\nu}}{\rho^2}\left(\nabla_{y}^2 A+ 8
{A^\prime}^2\right)\nonumber\\
 &=& {}^{(4)}\hat{R}_{\mu\nu} -
\hat{g}_{\mu\nu} \frac{e^{-8A(y)}}{8\rho^2}\nabla_{y}^2 e^{8A(y)}.
\end{eqnarray}
On the other hand, from the 10D Einstein equations $R_{AB}-(1/2)
g_{AB} R = 8\pi G_{10}\, T_{AB}$, we obtain
\begin{eqnarray}
{}^{(10)} R_{\mu\nu}&=& 8\pi G\Z{10} \left(T_{\mu\nu} -
\frac{1}{8}\,e^{2A} \hat{g}_{\mu\nu} \, T_C^C\right), \nonumber\\
{}^{(10)} R_\mu^\mu&=& 4\pi G\Z{10} \left( T_\mu^\mu -
T_m^m\right).
\end{eqnarray}
The above result shows that, with $\int \nabla_{y}^2 e^{8A(y)}=0$,
a de Sitter spacetime (with ${}^{(4)} \hat{R} >0$) cannot be
realized without sources of $T_{00}$ which violate the positive
energy condition, i.e. without violating the condition
$T_m^m-T_\mu^\mu\ge 0$ (see also the discussion below Eq.~(2.15)
in~\cite{GKP}). This is the no-go theorem of Maldacena and
Nunez~\cite{Malda-Nunez}.

The above discussion is special at least from two aspects. First,
it only covered the case $\gamma=2$, for which $e^{(2-\gamma)A} =
1$ for any choice of $A(y)$. Second, the condition on the warp
factor, i.e. $\int \nabla\Z{y}^2 e^{8A(y)}=0$ is `strict' and it
is not always satisfied, especially, in the presence of some brane
sources. Further, the 6d metric of the
form~(\ref{6d-flat})-(\ref{6d-negative}) was not sufficiently
general as it contained no free parameter that can be tuned or
fixed according to the choice of $\gamma$ in the warp factor. By
relaxing the condition like $\oint \nabla_y^2 e^{nA} =0$ (where
$n$ is some constant) or some other similar constraints one should
expect de Sitter solution to be easy to realize. Below we will
give a couple of explicit examples.

\section{An explicit model in $D=10$ dimensions}

Let us generalize the 6d metric in
Eq.~(\ref{6d-flat})-(\ref{6d-negative}) as follows
\begin{eqnarray}\label{6d-general-anz}
ds_6^2= {g}_{mn}^{(6)}(y)\, dy^m dy^n\equiv h(y) \,dy^2 + f(y)\,
ds\Z{X_5}^2,
\end{eqnarray}
where $h(y)$ and $f(y)$ are two arbitrary functions. $X_5$ can be
taken to be a usual five-sphere $S^5$ or some other compact
Einstein manifolds. One of the well motivated examples is the
Einstein-Sasaki space $(S^2\times S^2)\rtimes S^1$ with
metric~\cite{Candelas89a}
\begin{equation} ds^2\Z{X_5}=
\frac{1}{6} \left(e_{\theta_1}^2+e_{\phi_1}^2
+e_{\theta_2}^2+e_{\phi_2}^2\right)+ \frac{1}{9} \,e_\psi^2,
\end{equation}
where $e_{\theta_i}=d\theta_i$, $e_{\phi_i}=\sin\theta_i d\phi_i$
and $e_\psi\equiv d\psi +\cos\theta_1 d\phi_1+\cos\theta_2
d\phi_2$, $(\theta_1, \phi_1)$ and $(\theta_2, \phi_2)$ are
coordinates on each $S^2$ and $\psi$ is the coordinate of a $U(1)$
fiber. One could in principle start with six-dimensional deformed
conifold metrics without any conical singularities, such as
in~\cite{KS,Ish07a}, or a deformed six-sphere as considered
in~\cite{Mukohyama09k}, but we find the metric ansatz
(\ref{6d-general-anz}) sufficiently simple for the purpose of
solving 10D Einstein equations analytically. For the Ans\"atze
(\ref{10D-nogo}) and (\ref{6d-general-anz}), a straightforward
calculation gives
\begin{eqnarray}
&& {}^{(10)} R_{\mu\nu} (x,y)\nonumber ~~~~~~~~~~~~~~~~~~~~~~~ \\
&& \quad = {}^{(4)}\hat{R}_{\mu\nu}-\frac{\hat{g}_{\mu\nu}
e^{(2-\gamma)A}}{\rho^2 h} \left( \nabla_y^2 A + 2 (2+\gamma)
{A^\prime}^2\right)
\nonumber \\
&& \quad = {}^{(4)}\hat{R}_{\mu\nu}(x)-
\frac{\hat{g}_{\mu\nu}\,e^{(2-\gamma)A}}{\rho^2 h}
\nonumber\\
&& \quad \times \left[ A^{\prime\prime}+\frac{1}{2}\left(\frac{5
f^\prime}{f}-\frac{h^\prime}{h}\right)A^\prime
+2(2+\gamma){A^\prime}^2 \right].\label{10and4Curv}
\end{eqnarray}

\begin{eqnarray}
R_{yy} &=& - \frac{8+5\gamma}{2}
A^{\prime\prime}-2(2-\gamma){A^\prime}^2+\frac{(8+5\gamma)
h^\prime A^\prime}{4h}\nonumber\\
&{}& -\frac{5\gamma f^\prime A^\prime}{4f} +
\frac{5}{4}\left(\frac{{f^\prime}^2}{f^2}+\frac{h^\prime
f^\prime}{h f}-\frac{2 f^{\prime\prime}}{f}\right),
\end{eqnarray}
\begin{eqnarray}
{}^{(10)} R_{p q} &=& {}^{(6)} R_{p q} - \tilde{g}_{p q}
\Big(\frac{\gamma f A^{\prime\prime}}{2h}+\frac{\gamma(2+\gamma)f
{A^\prime}^2}{h}\nonumber \\
&{}& \quad +\frac{(8+9\gamma)f^\prime A^\prime}{4h} -\frac{\gamma
f h^\prime A^\prime}{4h^2}\Big),
\end{eqnarray}
where ${}^\prime\equiv \partial/\partial{y}$ and
\begin{equation}
{}^{(6)}{R}_{p q}=\left(4-\frac{3 {f^\prime}^2}{4 h f}
-\frac{f^{\prime\prime}}{2 h}+\frac{h^\prime f^\prime}{4
h^2}\right)\tilde{g}_{pq}.
\end{equation}
In the above $\tilde{g}_{pq}$ denote the metric components of the
base space $X_5$, which are independent of the $y$ coordinate.

\medskip

{\bf Example 1}. Assume that $h(y)=1$ and $f(y) \equiv \alpha\Z{1}
(y-y_0)^2$. The 6d metric takes the form
\begin{equation}\label{6dgen2}
ds_6^2 = {g}_{mn}^{(6)}(y)\, dy^m dy^n\equiv dy^2 + \alpha\Z{1}
(y-y_0)^2\, ds\Z{X_5}^2.
\end{equation}
With $y_0=0$, $y$ measures the radius of the base space $X_5$, so
the coordinate range is $0\le y\le \infty$. Especially, when
$\alpha_1\ne 1$, the metric~(\ref{6dgen2}) is singular at $y=y_0$.
This leads to an undesirable result that the warp factor
$e^{A(y)}$ vanishes at $y=y_0$. To see this one can solve the 10D
Einstein equations explicitly. The solution is given by
\begin{eqnarray}\label{sol-nonsin2}
e^{A(y)} &=& \left(\frac{3\mu^2 (2-\gamma)^2
(y-y\Z{0})^2}{32}\right)^{1/(2-\gamma)}, \nonumber \\
 \alpha\Z{1} &\equiv & \frac{(2-\gamma)^2}{8},
\end{eqnarray}
with the same scale factor as given in~(\ref{main-sol-scale1}).
This yields
\begin{equation}
\nabla_y^2 A=  \frac{8}{(2-\gamma)(y-y\Z{0})^2} + \frac{
4\delta(y)}{(2-\gamma)(y-y\Z{0})}.
\end{equation}
As is evident, this solution does not satisfy the constraint like
$\oint \nabla_y^2 A=0$ or $\oint \nabla_y^2 e^{n A(y)}=0$.

\medskip

{\bf Example 2}. Assume that $h(y)=\sinh^2 (y-y\Z{0})$ and $f(y)
\equiv \alpha\Z{1} \cosh^2(y-y_0)$. The 6d metric
is\begin{equation}\label{6dgen3} ds_6^2= \sinh^2{(y-y_0)}\, dy^2 +
\alpha\Z{1} \cosh^2{(y-y_0)}\, ds\Z{X_5}^2.
\end{equation}
With $\alpha\Z{1}\equiv (2-\gamma)^2/8$, the 10D Einstein
equations are explicitly solved for
\begin{equation}\label{sol-nonsin3}
e^{A(y)}= \left(\frac{3\mu^2 (2-\gamma)^2
\cosh^2{(y-y_0)}}{32}\right)^{1/(2-\gamma)}.
\end{equation}
The 10d metric solution is given by
\begin{eqnarray}\label{soln-in-u}
ds\Z{10}^2 &=& e^{2A(y)} \Bigg(ds\Z{4}^2 + \frac{32\rho^2
\tanh^2(y-y\Z{0})}{3\mu^2(2-\gamma)^2}\nonumber \\
&{}& \times \left(
dy^2+\frac{(2-\gamma)^2}{8}\,\coth^2(y-y\Z{0})\,
ds\Z{X_5}^2\right) \Bigg)\nonumber \\
&\propto & u^{4/(2-\gamma)} \Bigg(ds\Z{4}^2 +
\frac{32\rho^2}{3\mu^2 (2-\gamma)^2 u^2}\nonumber\\
&{}& \times \left( du^2+\frac{(2-\gamma)^2}{8}\,u^2
\,ds\Z{X_5}^2\right) \Bigg),
\end{eqnarray}
where $u \equiv \cosh (y-y\Z{0})\equiv \sqrt{32/3\mu^2
(2-\gamma)^2}\,e^{(2-\gamma)A/2}$. This gives
\begin{equation}
\nabla_y^2 A=  \frac{8\tanh^2 (y-y\Z{0})}{(2-\gamma)} + \frac{
4\tanh(y-y\Z{0})\,\delta(y)}{(2-\gamma)}.
\end{equation}
From Eq.~(\ref{10and4Curv}) we then obtain
\begin{eqnarray}
&& {}^{(10)} R_{\mu\nu}\nonumber \\
&&  = {}^{(4)}\hat{R}_{\mu\nu} - \frac{\hat{g}_{\mu\nu}}{\rho^2}
\left({3\mu^2} +\frac{3\mu^2
(2-\gamma)} {8}\,\coth{(y-y_0)}\,\delta(y)\right).\nonumber \\
\end{eqnarray}
The exact solution above violates the warp factor constraint like
$\int \nabla_y^2 e^{nA}=0$. Moreover, the 6d warped volume is not
constant. Rather it scales as $V_6^{\text w} \sim \int d\Omega_5
\int e^{(2+3\gamma)A} \sqrt{{g\Z{6}}}\, dy \sim \int
u^{(14+\gamma)/(2-\gamma)} du$. Although one can hope to get an
ideal situation with almost constant or slowly varying warped
factor, for instance, by invoking some non-perturbative effects
(as in KKLT model~\cite{KKLT}) or introducing certain
$\alpha^\prime$ corrections, the solution above is interesting is
the regard that the radius modulus, which scales as $|\tanh{y}|$,
is constant in the limit $y\to \infty$~\cite{Ish:09a}. Further,
unlike with some singular conifold metrics considered in the
literature, for instance~\cite{KS,Gibbons-Hull}, our solution is
regular everywhere.

\medskip
Coming back to the metric~(\ref{6dgen2}), it is not difficult to
see that the singularity of this metric at $y=y_0$ (especially,
when $\alpha\Z{1}\ne 1$) is just a coordinate artifact. To
quantify this, we may introduce a new coordinate ${z}$, which is
related to the usual coordinate $y$ via $y \propto e^{-\lambda
\,z}$ (where $\lambda$ is some constant). The 10D metric that
explicitly solves all of the Einstein equations and is consistent
with $Z_2$ symmetry ($z\to - z$) about the brane's position
($z=0$) is
\begin{eqnarray}\label{10d-soln}
ds\Z{10}^2 &=& e^{2A(z)} \left(ds_4^2 + \frac{8\rho^2}{3\mu^2\,
\ell^2}\, {dz}^2 + \frac{4\rho^2}{3\mu^2}\,
ds\Z{X_5}^2\right),\nonumber \\
A(z) &=& - \frac{|z|}{\ell} -\frac{A\Z{0}}{2}.
\end{eqnarray}
From this, we derive $\nabla_z^2 A=- (2/\ell)\delta(z)$ and hence
\begin{eqnarray}
{}^{(10)} R_{\mu\nu} = {}^{(4)}\hat{R}_{\mu\nu} - \frac{3\mu^2
\hat{g}_{\mu\nu}}{8\rho^2}  \left( -\,\frac{2}{\ell}\,\delta(z)+
\frac{8}{\ell^2} \right).
\end{eqnarray}
If we do not enforce the $Z_2$ symmetry, then the solution above
satisfies $\nabla_z^2 A =0$. In this case, a four-dimensional de
Sitter solution with ${}^{(4)}\hat{R}_{00}<0$ is still possible,
but there arises an important difference: since $z$ ranges from
$-\infty$ to $+\infty$, the 6d warped volume can be arbitrarily
large. Typically, $V_6^{{\text w}}\sim \int
e^{8A}\sqrt{\tilde{g}_6}\sim
\frac{64\sqrt{2}}{27}\,e^{-4A\Z{0}}\int d\Omega_5\int dz\,
e^{-(8/\ell) z}$, where $\int d\Omega_5=\frac{1}{108}\int
d(\cos\theta_1)d(\cos\theta_2)d\phi_1 d\phi_2 d\psi= 16\pi^3/27$.
To get a sensible result with an almost constant warped volume (or
slowly varying warp factor), we need to send $\ell \to \infty$ or
take $e^{-4 A\Z{0}}\to 0$.

The metric solution~(\ref{soln-in-u}) is already regular
everywhere, but we may introduce some brane sources at $y=y_0$ and
then solve the 10d Einstein equations with proper regularity
conditions at $y=y\Z{0}$. This was in fact done quite recently in
the second paper in~\cite{Ish:09a}, so in the following discussion
we only consider the metric solution~(\ref{10d-soln}).

To solve the Einstein equations at $z=0$, we shall write
\begin{equation}\label{10d-boundary}
G_A^B=  \tau_p\, P(g_A^B),
\end{equation}
where $P(g_A^B)$ is the pull-back of the spacetime to the world
volume of the $p$-brane ($3\le p\le 8$) with tension $\tau_p$. We
shall impose a $Z_2$ symmetry at the brane's position at $z=0$.
The warp factor will then have a discontinuity in its first
derivative, implying that $\partial |z|/\partial z=\text{sgn}(z)$
and $\partial^2|z|/\partial z^2= 2\delta(z)$. We then obtain
\begin{eqnarray}\label{10d-brane-sol}
{G_z^z}|\Z{z=0}=0, \quad G_M^N|\Z{z=0}=\frac{6 \mu^2 \ell
}{\rho^2}\,e^{A \Z{0}}\, \delta(0) \eta_M^N,
\end{eqnarray}
where $(M,N)=t,x_i, \theta_i, \phi_i, \psi$. We only consider the
simplest case that $p=8$, for which the brane extends to all of
the dimensions except along the $z$-direction and thus $P(g_A^B)=
\delta(z)/\sqrt{g_{zz}}$. Einstein's equations are satisfied at
$z=0$ when
\begin{equation}
\tau\Z{8} = \frac{4\sqrt{6}\,\mu}{\rho}.
\end{equation}

From Eq.~(\ref{10d-soln}) we derive
\begin{eqnarray}
M_{Pl}^2&=& \frac{M\Z{10}^8}{(2\pi)^6} \frac{32\sqrt{2}\,\rho^6
e^{-4 A\Z{0}}}{27\,\ell \mu^6}\int d\Omega_5
\int_{-\infty}^{\infty} dz\, e^{-8 |z|/\ell}\nonumber \\
&\approx & \frac{M\Z{10}^8}{\pi^3} \frac{16\sqrt{2}\,\rho^6 e^{-4
A\Z{0}}}{729\, \mu^6}.
\end{eqnarray}
In the above we made the approximation $\int_{-\infty}^{\infty}
dz\, e^{-8 |z|/\ell} \approx \ell/4$, which is reasonably good
when $z\to \infty$. Note that the warping becomes stronger away
from the brane at $z=0$.
This feature is similar to that in RS single brane model.

\subsection{Positive energy condition}

For several explicit solutions given above, inflationary cosmology
is possible without violating any energy condition in the full
D-dimensions. To quantify this, we can make an ansatz for the
stress-energy tensor of the form
\begin{equation}
T_{A}^{B}=  \tau_p P(g\Z{A}^B) +{\cal T}_{A}^{B}.
\end{equation} ${\cal T}_{A}^{B}$ represents the contribution
of bulk matter fields. In $D=5$ dimensions, and with $\gamma=2$,
we have
\begin{equation} R_5\,^5 - R_0\,^0=
\left[\frac{6{W^\prime}^2}{W^2} -\frac{3
W^{\prime\prime}}{W}-\rho^2\,\frac{3\ddot{a}}{a}\right]
\frac{1}{\rho^2 W^2},\nonumber
\end{equation}
where $W(z)= e^{-\mu|z|}$. In order not to violate the 5D null
energy condition (NEC) we require $ R_5\,^5 - R_0\,^0 \ge 0$. In
the simplest case that ${\cal T}_{AB}=0$, we find
$R_5\,^5-R_0\,^0= 6\mu/\rho^2>0 $ on the brane and
$R_5\,^5-R_0\,^0=0$ in the bulk. Similarly, for the 10D solution
given above, Eq.~(\ref{10d-soln}), we find
\begin{eqnarray}
&& \tilde{R}_{m n} \tilde{g}^{mn} - R_0\,^0 \nonumber\\
&& ~= \left[15-\frac{57 \ell^2}{8} \frac{{W^\prime}^2}{W^2}
-\frac{39 \ell^2}{8} \frac{
W^{\prime\prime}}{W}-\frac{\rho^2}{\mu^2}\,\frac{3\ddot{a}}{a}\right]
\frac{\mu^2}{\rho^2 W^{2}},\nonumber
\end{eqnarray}
where $W(z)= e^{-|z|/\ell}$. Again, $ \tilde{R}_{m n}
\tilde{g}^{mn} - R_0\,^0=0 $ in the bulk and $>0 $ on the brane.
There is no violation of any energy condition in the full
$D$-dimensional spacetime, and no violation of the null energy
condition in four dimensions. This result can be understood also
from the viewpoint that the NEC can be violated only by
introducing non-standard bulk matter fields (i.e. ${\cal
T}_0\,^0<0$) or by introducing negative tension branes or
orientifold planes~\cite{GKP} that may serve as sources of the NEC
violation in a subspace of the internal manifold.

The above explicit results may appear in conflict with a claim
made in~\cite{Wesley:08fg}, where it was argued that to get a
four-dimensional de Sitter space solution one may have to violate
the five- and higher-dimensional null energy conditions or allow a
time-dependent Newton's constant or even both. There is perhaps no
contradiction here, since the discussion in~\cite{Wesley:08fg} may
apply only to a particular model with physically compact extra
spaces, supplemented with additional constraints on the warp
factor. String theory can, of course, accommodate some NEC
violating objects, such as negative tension branes and orientifold
planes~\cite{GKP}, but in our view such objects are not necessary
just to get an accelerating universe from higher-dimensional
Einstein's theory.

\medskip

In conclusion, we have proposed an alternative scenario to
conventional explanation to cosmic acceleration, by embedding a
four-dimensional de Sitter spacetime into higher-dimensional
spacetimes. We have shown the existence of inflationary cosmology
in a wide class of metrics, obtaining explicit cosmological
solutions both in five- and ten-dimensions, which do not violate
the higher-dimensional positive energy condition. In $D=10$
dimensions, our solutions correspond to the dimensional reduction
to four dimensions of $d=10$ supergravity (with zero flux), where
the spacetime is a warped product of a four-dimensional de Sitter
space dS$_4$ and a six-dimensional Einstein space $E_6$ (with
arbitrary curvature). We only took into account contributions from
brane sources and metric flux (arising as a nontrivial effect of
the internal curvature), so the present construction may be viewed
as a local model. The no-go arguments for de Sitter solutions as
simple as the one given for classical supergravities with
fluxes~\cite{deWit86a,Malda-Nunez} or the one for string flux
compactifications~\cite{GKP} may not be applied to our examples
because we considered less symmetric spacetimes with arbitrary
curvatures, and also relaxed some of the conditions imposed on the
warp factor.

One of the remarkable features of our model is that the brane
tension is induced not by a bulk cosmological constant but by the
curvature related to the expansion of the physical $3+1$
spacetime, which appears to vanish only in the limit where the
scale factor becomes a constant. The universe accelerates when
$\mu>0$, giving rise to a nontrivial warp factor, and the brane
tension becomes positive. There is no static limit of our
solutions: the scale factor (of the universe) becomes a constant
(in a spatially flat FRW universe) only when $\mu= 0$, but in this
case the warp factor is also constant and the brane tension
vanishes. In the generic situation with a $Z_2$ symmetry about the
brane's position, and with a nonzero Hubble parameter, the
four-dimensional Newton's constant is effectively finite.

Within our model the four-dimensional effective cosmological
constant is given by $\Lambda\Z{4}=6 \mu^2/\rho^2$, to leading
order. It is clearly determined in terms of two length scales: one
is a scale associated with the size of extra dimensions or the
compactification scale, $\rho$, and the other is a scale
associated with the curvature related to the expansion of the
physical three spaces, which also determines the slope of warp
factor. This could just be due to a general fact that the warp
factor relates energy scales on compactified spaces to those in
$3+1$ spacetime.

We conclude with the following remark. Recently, important steps
have been taken in the literature toward investigating minimal de
Sitter solutions in type IIA and IIB string
theory~\cite{Silverstein,Shiu-etal,Douglas:09}. In most of these
works, one adopts a common notion that the low energy effective
potential (and hence the gravitational vacuum energy density of
our universe) is a sum of the effects from different regions of
the internal manifold, supergravity fluxes and effects of
localized sources like branes and orientifold planes, and then
check certain conditions under which the effective potential
allows one or more metastable de Sitter minima. Though this
exercise seems reasonable from a viewpoint of effective field
theory, it would be more beneficial to know some explicit
cosmological solutions at least within some workable models; we
leave the analysis of this nature, specific to our metric choices,
for subsequent work.

\medskip

{Acknowledgements:} I am grateful to Bruce Bassett, Robert
Brandenberger, Rong-Gen Cai, Xingang Chen, Hassan Firouzjahi, Juan
Maldacena and, especially, Nobu Ohta for valuable conversations
and helpful comments. This research was supported by the New
Zealand Foundation for Research, Science and Technology Grant No.
E5229 and also by Elizabeth EE Dalton Grant No. 5393.


\end{document}